# Probability Bracket Notation, Term Vector Space, Concept Fock Space and Induced Probabilistic IR Models

Xing M. Wang
Sherman Visual Lab, Sunnyvale, CA 94085, USA

## Table of Contents

## Abstract

After a brief introduction to *Probability Bracket Notation* (*PBN*) for discrete random variables in time-independent probability spaces, we apply both *PBN* and Dirac notation to investigate probabilistic modeling for information retrieval (IR). We derive the expressions of relevance of document to query (RDQ) for various probabilistic models, induced by *Term Vector Space* (TVS) and by *Concept Fock Space* (CFS). The inference network model (INM) formula is symmetric and can be used to evaluate relevance of document to document (RDD); the CFS-induced models contain ingredients of all three classical IR models. The relevance formulas are tested and compared on different scenarios against a famous textbook example.

## 1. Introduction: From Hilbert Space to Probability Space

Dirac notation is a very powerful tool to manipulate vectors in Hilbert spaces [1]. It has been widely used in Quantum Mechanics (QM) and Quantum Field Theories.

Suppose we have a time-independent discrete observable $\hat{H}$ (a Hermitian operator). In Dirac notation, its complete set of eigenvectors has the following properties:

$$\hat{H}\,|\,\varepsilon_j\rangle = \varepsilon_j\,|\,\varepsilon_j\rangle, \quad \langle\varepsilon_j\,|\,\hat{H} = \langle\varepsilon_j\,|\,\varepsilon_j, \quad \langle\varepsilon_i\,|\,\varepsilon_j\rangle = \delta_{ij}, \quad \sum_i |\,\varepsilon_i\rangle\langle\varepsilon_i\,| = \hat{I} \tag{1.1}$$

The expectation value of $\hat{H}$ in a given normalized system state $|\,\psi\rangle$ can be written as:

$$\bar{H} \equiv \langle\psi|\hat{H}|\psi\rangle = \langle\psi|\hat{H}\hat{I}|\psi\rangle = \sum_i \langle\psi|\hat{H}|\varepsilon_i\rangle\langle\varepsilon_i|\psi\rangle$$
$$= \sum_i \varepsilon_i \langle\psi|\varepsilon_i\rangle\langle\varepsilon_i|\psi\rangle = \sum_i \varepsilon_i \left|\langle\varepsilon_i|\psi\rangle\right|^2 = \sum_i \varepsilon_i P(\varepsilon_i) \quad (1.2)$$

Hence, the probability of observable $\hat{H}$ having exact value $\varepsilon_i$ in state $|\psi\rangle$ is given by:

$$P(\varepsilon_i) = \left|\langle\varepsilon_i|\psi\rangle\right|^2 \quad (1.3)$$

The unit operator in Eq. (1.1) can be "derived" by using the normalization property:

$$1 = \sum_i P(\varepsilon_i) = \sum_i \left|\langle\varepsilon_i|\psi\rangle\right|^2 = \sum_i \langle\psi|\varepsilon_i\rangle\langle\varepsilon_i|\psi\rangle = \langle\psi|\left\{\sum_i |\varepsilon_i\rangle\langle\varepsilon_i|\right\}|\psi\rangle \quad (1.4a)$$
$$1 = \langle\psi|\psi\rangle \quad (1.4b)$$

Compare (1.4a) with (1.4b), we "derive" the unit operator in Hilbert space:

$$\sum_i |\varepsilon_i\rangle\langle\varepsilon_i| = \hat{I} \quad (1.4c)$$

Now let us go to probability space. All possible outcomes of the random variable $H$ form a sample space Ω, and by definition of conditional probability, we can write:

$$P(\varepsilon_i) = P(\varepsilon_i|\Omega) = \left|\langle\varepsilon_i|\psi\rangle\right|^2, \quad P(\Omega|\varepsilon_i) = 1, \quad P(\Omega|\Omega) = 1 \quad (1.5)$$

We call it the ***induced probability space*** from Hilbert space (1.1). Moreover:

$$1 = \sum_i P(\varepsilon_i) = \sum_i P(\varepsilon_i|\Omega) = \sum_i P(\Omega|\varepsilon_i)P(\varepsilon_i|\Omega) \underset{(1.5)}{=} P(\Omega|\left\{\sum_i |\varepsilon_i)P(\varepsilon_i|\right\}|\Omega) \quad (1.6)$$

Compare (1.6) with (1.5), we "derive" the unitary operator in probability space:

$$\sum_i |\varepsilon_i)P(\varepsilon_i| = I \quad (1.7)$$

Let us also propose that:

$$H|\varepsilon_j) = \varepsilon_j|\varepsilon_j) \quad (1.8)$$

Then the expectation value of variable $H$ in a given probability space can be written as:

$$E[H] = P(\Omega | H | \Omega) = P(\Omega | H I \ | \Omega) \underset{(1.7)}{=} \sum_i P(\Omega | H | \varepsilon_i) P(\varepsilon_i | \Omega)$$
$$\underset{(1.8)}{=} \sum_i \varepsilon_i P(\Omega | \varepsilon_i) P(\varepsilon_i | \Omega) \underset{(1.5)}{=} \sum_i \varepsilon_i P(\varepsilon_i) \underset{(1.2)}{=} \langle \psi | \hat{H} | \psi \rangle = \bar{H} \tag{1.9}$$

We see that Dirac notation can be naturally extended to probability space, which we call the *Probability Bracket Notation* (*PBN*) [3]. It is a new set of symbols for probability modeling.

In next section, we will introduce it for time-independent discrete random variables. Then we will apply both *PBN* and Dirac notation to discuss various *probabilistic IR models*, *induced* by Term Vector Space (TVS) and Concept Fock Space (CFS). The symmetric INM model is used to derive RDQ and RDD formulas, which also contain the ingredients of all three classical IR models. We will test and compare our models by using a famous textbook example on a mixture of scenarios.

## 2. PBN for Discrete Random Variables and Bayesian Inference

**Discrete random variable**: We define a probability space (Ω, *X*, *P*) of a discrete random variable (observable) *X* as follows: the set of all elementary events ω, associated with a discrete random variable *X*, is the sample space Ω, and

$$\text{For } \forall \omega_i \in \Omega, X(\omega_i) = x_i \in \Re, \quad P: \omega_i \mapsto P(\omega_i) = m(\omega_i) \ge 0, \sum_i P(\omega_i) = 1 \tag{2.1}$$

**Proposition 1** (*Probability event-bra and evidence-ket)***:** Let $A \subseteq \Omega$ and $B \subseteq \Omega$,

1. The symbol $P(A|$ represents a probability event bra, or *P*-bra;
2. The symbol $|B)$ represents a probability evidence ket, or *P*-ket.

**Proposition 2** (*Probability Event-Evidence Bracket*): The *conditional probability* of event *A* given evidence *B* in the sample space Ω is also called *P-bracket*, and it can be split into a *P*-bra and a *P*-ket, similar to a Dirac bracket:

$$P(A | B) \equiv \frac{P(A \cap B)}{P(B)} = \frac{|A \cap B|}{|B|}, \text{ if } 0 < \frac{|B|}{|\Omega|} \le 1 \tag{2.2a}$$
$$P\text{-braket} \quad P(A | B) \ \Rightarrow \ P\text{-bra}: \ P(A|, \quad \text{P-ket}: \ |B) \tag{2.2b}$$
$$P(A | B) = P(A | I | B), \quad \text{where } I \text{ is a unit operator} \tag{2.2c}$$

By definition, the *P*-bracket has the following properties for discrete sample space Ω:

$$P(A | B) = 1 \quad \text{if } \Omega \supseteq A \supseteq B \supset \varnothing \tag{2.3}$$
$$P(A | B) = 0 \quad \text{if } A \subset \Omega, B \subset \Omega \text{ and } A \cap B = \varnothing \tag{2.4}$$

We can see that *P-bracket is not the inner product* of two vectors. For any event $E \subseteq \Omega$, the absolute probability *P(E)* now can be written as:

$$P(E) = P(E \mid \Omega) \qquad (2.5)$$

Here $|\Omega)$ is called the *system P-ket*. The *P*-bracket defined in (2.2a) now becomes:

$$P(A \mid B) = \frac{P(A \cap B)}{P(B)} = \frac{P(A \cap B \mid \Omega)}{P(B \mid \Omega)} \qquad (2.6)$$

Properties in Eq. (2.3-4) can be easily verified by using this definition.

The *Bayes formula* (see [2], §2.1) now can be expressed as:

$$P(A \mid B) = \frac{P(B \mid A)P(A)}{P(B)} = \frac{P(B \mid A)P(A \mid \Omega)}{P(B \mid \Omega)} \qquad (2.7)$$

The set of all elementary events in Ω forms a complete mutually disjoint basis:

$$\bigcup_{\omega_i \in \Omega} \omega_i = \Omega, \quad \omega_i \cap \omega_j = \delta_{ij}\, \omega_i, \quad \sum_i P(\omega_i) = 1 \qquad (2.8)$$

**Proposition 3** (*Discrete P-Basis and Unit Operator*): Using Eq. (2.1-4), Eq. (1.8) and (2.8), we have following properties for *basis* elements in (Ω, *X*, *P*):

$$X \mid \omega_j) = x_j \mid \omega_j), \quad P(\omega_j \mid X = P(\omega_j \mid x_j, \quad P(\Omega \mid \omega_j) = 1, \quad P(\omega_i \mid \Omega) = P(\omega_i) \qquad (2.9)$$

The complete mutually-disjoint events in (2.8) form a *probability sample basis* (or *p-basis*) and a unit (or identity) operator:

$$P(\omega_i \mid \omega_j) = \delta_{ij}, \quad \sum_{\omega \in \Omega} \mid \omega)P(\omega \mid = \sum_{i=1} \mid \omega_i)P(\omega_i \mid = I. \qquad (2.10)$$

The system *P*-ket, denoted by $|\Omega)$, now can be right-expanded as:

$$\mid \Omega) = I \mid \Omega) = \sum_i \mid \omega_i)P(\omega_i \mid \Omega) = \sum_i P(\omega_i) \mid \omega_i) \qquad (2.11)$$

While for the system *P*-bra, denoted by $P(\Omega \mid$, has its left-expansion as:

$$P(\Omega \mid = P(\Omega \mid I = \sum_i P(\Omega \mid \omega_i)P(\omega_i \mid \underset{(2.9)}{=} \sum_i P(\omega_i \mid \qquad (2.12)$$

The two expansions are quite different, and $(\Omega \mid \neq [\mid \Omega)]^\dagger$. But their *P*-bracket is consistent with the requirement of normalization:

$$1 = P(\Omega) \equiv P(\Omega | \Omega) = \sum_{i,j=1} P(\omega_i | P(\omega_j) | \omega_j) = \sum_{i,j=1} P(\omega_j) \delta_{ij} = \sum_{i=1} P(\omega_i) \qquad (2.13)$$

**Proposition 4** (*Expectation Value*): The expected value of the observable $X$ in **Ω** now can be expressed as:

$$\langle X \rangle \equiv \bar{X} \equiv E(X) = P(\Omega | X | \Omega) = \sum_{x \in \Omega} P(\Omega | X | x) P(x | \Omega) = \sum_{x \in \Omega} x P(x) \qquad (2.14)$$

If *F(X)* is a continuous function of observable X, then it is easy to show that:

$$\langle F(X) \rangle \equiv E(F(X)) \equiv P(\Omega | F(X) | \Omega) = \sum_{x \in \Omega} F(x) P(x) \qquad (2.15)$$

***Joint random variable***: Let $N_1, N_2, \dots N_n$ be random variables associated with a probability space. Suppose that the sample space (i.e., the set of possible outcomes) of $N_i$ is the set $\Omega_i$. Then the *joint random variable* (or *random vector*) is denoted as $\vec{N} = (N_1, N_2, \dots, N_n)$. The sample space of $\vec{N}$ is the Cartesian product of the $\Omega_i$'s:

$$\Omega = \Omega_1 \otimes \Omega_2 \otimes \dots \otimes \Omega_n \qquad (2.16)$$

**Proposition 5** (*Factor Kets*): The sample space of joint variable $\vec{N}$ now can be written as:

$$| \Omega) = \prod_{i=1}^{n} | \Omega_i) \qquad (2.17)$$

The factor *P*-ket$| \Omega_i)$ has the following properties:

$$P(\Omega_i | \Omega_i) = 1, \quad | \Omega_i) | \Omega_j) = | \Omega_j) | \Omega_i), \quad P(\Omega_i | \times P(\Omega_j | = P(\Omega_j | \times P(\Omega_i | \qquad (2.18)$$

As an example, in ***Fock space***, we have the following basis from the *occupation numbers*

$$N_i | \vec{n}) = n_i | \vec{n}), \quad P(\vec{n} | \vec{n}') = \delta_{\vec{n},\vec{n}'} = \prod_i \delta_{n_i, n'_i} \quad \sum_{\vec{n}} | \vec{n}) P(\vec{n} | = I \qquad (2.19)$$

The expectation value of an occupation number now is given by:

$$\langle N_i \rangle \equiv P(\Omega | N_i | \Omega) = P(\Omega_i | N_i | \Omega_i) = \sum_k k P(k | \Omega_i) \qquad (2.20)$$

If elements in set $\{X_1, X_2, \dots X_k)$ belong to different factor spaces, then they are ***mutually independent.*** Assuming $\varnothing \subset H \subseteq \Omega$, we have [2],

$$P(X_1 \wedge X_2 \wedge \dots \wedge X_k | H) = \prod_{i=1}^{k} P(X_k | H) \qquad (2.21)$$

***Bayesian Inference* and Unitary Operators**: Now we want to use a simple example of Bayesian inference [11] to explain the usability of unit operators in probability modeling.

Suppose there are two full bowls of cookies. Bowl #1 ($H_1$) has 10 chocolate chip (F) and 30 plain cookies (E), while bowl #2 ($H_2$) has 20 of each. Our friend Fred picks a bowl at random, and then picks a cookie at random. How probable is it that Fred picked a plain cookie out of bowl #1, or $P(H_1| E)$? The bowls are identical, thus $P(H_1) = P(H_2) = 0.5$. Also we know that:

$$P(E \mid H_1) = 30 / 40 = 0.75 \text{ and } P(E \mid H_2) = 20 / 40 = 0.5.$$

One can start with Bayes' formula:

$$P(H_1 \mid E) = \frac{P(E \mid H_1)P(H_1)}{P(E)} \tag{2.22}$$

To evaluate $P(E) = P(E \mid \Omega)$, we can use following unit operator:

$$\Omega = H_1 \cup H_2, \quad H_1 \cap H_2 = \varnothing, \quad \therefore I = |H_1)P(H_1| + |H_2)P(H_2|$$

$$\therefore P(E \mid \Omega) = P(E \mid I \mid \Omega) = P(E \mid H_1)P(H_1 \mid \Omega) + P(E \mid H_2)P(H_2 \mid \Omega)$$

$$\therefore P(E) = P(E \mid H_1)P(H_1) + P(E \mid H_2)P(H_2) = 0.75 \times 0.5 + 0.5 \times 0.5 = 0.625$$

$$\therefore P(H_1 \mid E) = \frac{P(E \mid H_1)P(H_1)}{P(E)} = \frac{0.75 \times 0.5}{0.625} = 0.6 \tag{2.23}$$

One can also directly evaluate $P(H_1 \mid E)$ as follows:

$$\because H_1 = E_1 \cup F_1, \quad H_2 = E_2 \cup F_2, \quad E_i \cap F_j = \varnothing, \quad E_1 \cap E_2 = \varnothing, \quad F_1 \cap F_2 = \varnothing$$

$$\therefore I = |E_1)P(E_1| + |E_2)P(E_2| + |F_1)P(F_1| + |F_2)P(F_2|$$

$$\therefore P(H_1 \mid E) = P(H_1 \mid E_1)P(E_1 \mid E) + P(H_1 \mid E_2)P(E_2 \mid E) = P(E_1 \mid E)$$

$$\therefore P(H_1 \mid E) = \frac{|E_1|}{|E|} = \frac{30}{30+20} = 0.6 \tag{2.24}$$

It is identical to what we get from Eq. (2.23).

## 3. Probabilistic Models Induced by Term Vector Space

In reference [4], we used Dirac notation to describe the three classical IR models: vector space model, Boolean model and probabilistic model in their classical ways [5, 6]. Then we introduced Fock space as a unified way to represent the above models. But, as readers may have noticed, the probabilistic model was not well integrated in this framework. The main reason was that we could not directly apply Dirac notation to probability spaces.

This actually was one major motivation of our proposal of *PBN* [3]. Now we will show how *PBN* would help us to simplify and unify expressions of probabilistic IR models.

***Term Vector Space*** (TVS): First, let us consider the $t$-dimensional vector space spanned by indexed terms, a set $K = \{k_1, k_2, \ldots, k_t\}$. The terms are assumed to be mutually disjoint, so the Hilbert space has the following basis [4]:

$$\langle k_i \mid k_j \rangle = \delta_{ij}, \quad \sum_{i=1}^{t} | k_i \rangle \langle k_i | = I \tag{3.1}$$

In this space, query $q$ and document $d_\mu$ ($\mu$=1, 2…$N$) can be naturally expanded as:

$$| q \rangle = I | q \rangle = \sum_{i=1}^{t} | k_i \rangle \langle k_i | q \rangle = \sum_{i=1}^{t} w_{q,i} | k_i \rangle; \quad | d_\mu \rangle = \sum_{i=1}^{t} | k_i \rangle \langle k_i | d_\mu \rangle = \sum_{i=1}^{t} w_{\mu.i} | k_i \rangle. \tag{3.2}$$

Here, $w_{q,i}$ and $w_{\mu,i}$ are called ***term weights*** of a query or a document with respect to the $i^{\text{th}}$ term [5, 6]. As vectors in Hilbert space, $d_\mu$ and $q$ are normalized as in Eq. (1.4):

$$\vec{d}_\mu \cdot \vec{d}_\mu \equiv \langle d_\mu | d_\mu \rangle = \sum_{i=1}^{t} w_{\mu,i}^2 = 1; \quad \vec{q} \cdot \vec{q} \equiv \langle q | q \rangle = \sum_{i=1}^{t} w_{q,i}^2 = 1 \tag{3.3}$$

Mapping Eq. (3.1) to the induced sample space $\Omega_V$, the set of all elements of $K$, we have the following $P$-basis, as a special case of Eq. (2.10),

$$P(k_i | k_j) = \delta_{ij}, \quad \sum_{i=1}^{t} | k_i ) P(k_i | = I \tag{3.4}$$

Because each document or query represents a normalized state vector in the Hilbert space, from Eq. (1.3), we have the following induced probability distribution functions:

$$P(k_i | d_\mu) = \langle k_i | d_\mu \rangle^2 = w_{\mu,i}^2, \quad P(k_i | q) = \langle k_i | q \rangle^2 = w_{q,i}^2 \tag{3.5}$$

Term weights are functions of ***term frequencies*** [5, 6]. We will introduce some other weight formulas (WF) later in this section, but the most natural way is to assume documents and query as normalized vectors in a $t$-dimensional Euclidian space:

$$\text{WF-1:} \quad w_{\mu,i} = \frac{tf_{\mu,i}}{\sqrt{\sum_{j=1}^{t} tf_{\mu,j}^2}} \equiv \frac{tf_{\mu,i}}{\sqrt{W_\mu}}, \qquad w_{q,i} = \frac{tf_{q,i}}{\sqrt{W_\mu}} \tag{3.6}$$

Now we are ready to derive formulas for TVS-induced probabilistic models.

***Inference Network Model*** (INM, §2.8.2 of [5]): In this model, the relevance of a document $d_\mu$ with respect to a query $q$ (RDQ) is given by $P(q \wedge d_\mu)$. Based on Eq. (2.6), we have two ways to evaluate RDQ:

$$P(q \wedge d_\mu) = P(q \mid d_\mu) P(d_\mu) \tag{3.7a}$$

$$P(d_\mu \wedge q) = P(d_\mu \mid q) P(q) \tag{3.7b}$$

They should produce the same symmetric result. Indeed, from (3.7a):

$$P(q \wedge d_\mu) \underset{(3.7a)}{=} P(q \mid d_\mu) P(d_\mu) = P(q \mid I \mid d_\mu) P(d_\mu) \underset{(3.4)}{=} P(d_\mu) \sum\nolimits_{i=1}^{t} P(q \mid k_i) P(k_i \mid d_\mu)$$

From TVS, we have $P(k_i \mid d_\mu) = w_{\mu.i}^2$ and $P(k_i \mid q) = w_{q,i}^2$. But $P(q \mid k_i)$ is not given by TVS. So we need Bayes formula (2.7), or ***Bayesian Inference*** to get it:

$$P(q \mid k_i) \underset{(2.7)}{=} \frac{P(k_i \mid q) P(q)}{P(k_i)}$$

Therefore:

**INM RDQ**: $P(q \wedge d_\mu) = P(d_\mu \wedge q) = P(q) P(d_\mu) \sum\nolimits_{i=1}^{t} \frac{1}{P(k_i)} P(k_i \mid q) P(k_i \mid d_\mu)$ (3.8)

So far, Eq. (3.8) is an exact expression, derived from first principals. The conditional probabilities $P(k_i \mid d_\mu)$ and $P(k_i \mid q)$ are given by the induced formula, Eq. (3.5). But there are no induced expressions for *absolute probabilities* $P(d_\mu)$, $P(q)$ and $P(k_i)$ (**APDQK**). They can be evaluated in various ways. Here are two of them:

**APDQK- I:** According to the ***tf-idf ranking strategies*** (pp. 54-55, [5]):

$$P(d_\mu) = P(d_\mu \mid \Omega_V) \propto \frac{1}{|d_\mu|} \propto \frac{1}{\sum_{i=1}^{t} tf_{\mu,i}}, \quad P(q) = P(q \mid \Omega_V) \propto \frac{1}{|q|} \propto \frac{1}{\sum_{i=1}^{t} tf_{q,i}} \tag{3.9a}$$

Furthermore, we may presuppose that the probability to find a term $k_i$ is proportional to $N_i$, the number of documents containing it. To avoid possible singularity, we assume:

$$P(k_i) = P(k_i \mid \Omega_V) \propto \frac{N_i + 1}{N + 1} \tag{3.9b}$$

**APDQK- II:** By using *Boolean Term Occupancy Number Representation* (BTON) in *Boolean Term Vector Space* (BT-Space) $n_{\mu,i}$ and $n_{q,i}$ (see §1.3 of [4]):

$$n_{\mu,i} = \begin{cases} 1, & \text{if } k_i \in d_\mu \\ 0, & \text{otherwise} \end{cases}, \qquad n_{q,i} = \begin{cases} 1, & \text{if } k_i \in q \\ 0, & \text{otherwise} \end{cases}, \qquad N_i = \sum_{\mu=1}^{N} n_{\mu,i} \tag{3.10a}$$

As concepts, $d_\mu$ and $q$ can be expressed as subsets of BT-Space $\Omega_V$ by using TB-ONR:

$$d_\mu = \bigcup_{i=1}^{t}(n_{\mu,i} \cdot k_i), \quad q = \bigcup_{i=1}^{t}(n_{q,i} \cdot k_i), \quad \Omega_V = \bigcup_{i=1}^{t}(k_i)$$

Then, by definition, we can obtain following expressions:

$$P(q) = \frac{|q|}{|\Omega_V|} = \frac{\sum_{i=1}^{t} n_{q,i}}{t}, \quad P(d_\mu) = \frac{|d_\mu|}{|\Omega_V|} = \frac{\sum_{i=1}^{t} n_{\mu,i}}{t}, \quad P(k_i) = \frac{|k_i|}{|\Omega_V|} = \frac{1}{t} \tag{3.10b}$$

Now we have the RDQ expression for TVS-INM based on term frequencies:

$$\text{INM:} \; P(q \wedge d_\mu) = C \cdot P(q) P(d_\mu) \sum_{i=1}^{t} \frac{w_{q,i}^2 \times w_{\mu,i}^2}{P(k_i)} \tag{3.11}$$

In ***Belief Network Model*** (BNM, §2.8.3 of [5] and [8]), the ranking of document $d_\mu$ with respect to query $q$ is given by $P(d_\mu \mid q)$. We can easily derive its expression from (3.11):

$$\text{BNM:} \; P(d_\mu \mid q) = \frac{P(d_\mu \wedge q)}{P(q)} \underset{(3.11)}{=} C \cdot P(d_\mu) \sum_{i=1}^{t} \frac{w_{q,i}^2 \times w_{\mu,i}^2}{P(k_i)} \tag{3.12}$$

Because only a q-dependent factor is removed, Eq. (3.12) will produce the same results as Eq. (3.11) when calculating RDQ. Therefore, we will not use it in our test.

According to ***Ponte and Croft*** (P&C, [9]), conditional probability $P(q \mid d_\mu)$ is adopted as the rank of document $d_\mu$ with respect to query $q$. From (3.11), its expression reads:

$$\text{P\&C:} \; P(q \mid d_\mu) = \frac{P(q \wedge d_\mu)}{P(d_\mu)} \underset{(3.11)}{=} C \cdot P(q) \sum_{i=1}^{t} \frac{w_{q,i}^2 \times w_{\mu,i}^2}{P(k_i)} \tag{3.13}$$

Furthermore, because Eq. (3.11) is symmetric, it can be used to calculate RDD:

$$\text{INM:} \; P(d_\mu \wedge d_\nu) = C \cdot P(d_\mu) P(d_\nu) \sum_{i=1}^{t} \frac{w_{\mu,i}^2 \times w_{\nu,i}^2}{P(k_i)} \tag{3.14}$$

***Reduction to vector space model*** (VSM): If we *forget* the relation in Eq. (3.5) and assume that $P(q\,|\,d_\mu)$ and $P(d_\mu\,|\,q)$ are evaluated as inner vector products $\langle q\,|\,d_\mu\rangle$ and $\langle d_\mu\,|\,q\rangle$, then they both lead to the same result as in classical VSM (§2.5.3 of [5]):

$$\text{VSM-RDQ: } P(d_\mu\,|\,q) = \langle d_\mu\,|\,q\rangle = P(q\,|\,d_\mu) = \sum\nolimits_{i=1}^{t} w_{\mu,i} w_{q,i} \tag{3.15}$$

$$\text{VSM-RDD: } P(d_\mu\,|\,d_\nu) = \langle d_\mu\,|\,d_\nu\rangle = \sum\nolimits_{i=1}^{t} w_{\mu,i} w_{\nu,i} \tag{3.16}$$

***More on weight formulas*** (***FS***): In classical vector models, in addition to Eq. (3.6), term weights may be calculated in other ways [5, 6]. They may use following definitions:

$$\text{Inverse document frequency (IDF): } \quad idf_i = \log\frac{N}{N_i} \tag{3.17}$$

$$\text{Normalized frequency: } \quad f_{\mu,i} \equiv \frac{tf_{\mu,i}}{\max_k tf_{\mu,k}}; \quad f_{q,i} \equiv \frac{tf_{q,i}}{\max_k tf_{q,k}} \tag{3.18}$$

Ref [6] (pp.14-18) provides us with two more ways to calculate weights. The simple one (after normalization) is:

$$\text{WF-2: } w_{\mu,i} = \frac{tf_{\mu,i}\cdot idf_i}{\sqrt{\sum_{i=1}^{t}\left(tf_{\mu,i}\cdot idf_i\right)^2}}, \quad w_{q,i} = \frac{tf_{q,i}\cdot idf_i}{\sqrt{\sum_{i=1}^{t}\left(tf_{\mu,i}\cdot idf_i\right)^2}} \tag{3.19}$$

A better one is from pp. 17-18 of [6], by replacing $\log(tf)+1$ with $\log(tf+1)$:

$$\text{WF-3: } w_{\mu,i} = \frac{\log(1+tf_{\mu,i})\cdot idf_i}{\sqrt{\sum_{i=1}^{t}\left[\log(1+tf_{\mu,i})\cdot idf_i\right]^2}}, \quad w_{q,i} = \frac{\log(1+tf_{q,i})}{\sqrt{\sum_{i=1}^{t}\left[\log(1+tf_{\mu,i})\cdot idf_i\right]^2}} \tag{3.20}$$

On the other hand, the best known *term-weighing schemes* use the following modified formulas (see §2.5.3 of [5]):

$$\text{WF-4: } w_{\mu,i} = \frac{f_{\mu,i}\cdot idf}{\sqrt{\sum_{i=1}^{t}(f_{\mu,i}\cdot idf_i)^2}}, \quad w_{q,i} = \frac{\left(1+f_{q,i}\right)\cdot idf}{\sqrt{\sum_{i=1}^{t}\left((1+f_{q,i})\cdot idf_i\right)^2}} \tag{3.21}$$

We will use all four weight formulas, Eq. (3.6) and Eq. (3.19-21), to test the relevance expressions of our models. The weights for the GF-example (see App A) are listed in App B. Our test results (see §5) show that weights from Eq. (3.19) seem to produce the best ranking, while results using weights from (3.21) seem not to be consistent with others.

## 4. Models Induced by Concept Fock Space and Unified IR Modeling

Based on our discussion in [4], if we have $t$ indexed terms $K = \{k_1, k_2, \ldots, k_t\}$, then we have $2^t$ elementary disjoint concepts, the same number of outcomes as we toss coin $t$-times. These elementary disjoint concepts form the sample space $\Omega$ of the ***probabilistic concept space*** (PCS) [7]. The $2^t$-dimensional basis for PCS has following properties:

$$P(\vec{n} \mid \vec{n}') = \delta_{\vec{n},\vec{n}'} = \prod_{i=1}^{t} \delta_{n_i,n'_i} \qquad \sum_{\vec{n}} |\vec{n}) P(\vec{n}| = \hat{I} \tag{4.1a}$$

But, if we use this basis, then the flowing conditional probability almost always vanishes:

$$P(\vec{q} \mid \vec{d}_\mu) = \sum_{\vec{n}} P(\vec{q} \mid \vec{n}) P(\vec{n} \mid \vec{d}_\mu) = \sum_{\vec{n}} \delta_{\vec{n}_q, \vec{n}_\mu} = 0, \quad \text{if } \exists i : n_{q,i} \neq n_{\mu,i} \tag{4.1b}$$

That means, it vanishes unless the query and the document contain the same terms.

***Concept Fock space*** (CFS): Now let us look at the basis in the *Occupation Number Representation* ([1], page 566) as a fermions Fock space (or *Boolean Fock space* [4]):

$$|\vec{n}\rangle = |n_1, n_2, \ldots n_t\rangle = \sum_{i=1}^{t} n_i |\vec{n}_i\rangle, \quad |\vec{n}_1\rangle = |1,0,...,0\rangle, ..., |\vec{n}_t\rangle = |0,0,...,1\rangle \tag{4.2a}$$

$$\hat{n}_i |\vec{n}\rangle = n_i |\vec{n}\rangle, \quad n_i \in \{0,1\}, \quad 1 \le i \le t \tag{4.2b}$$

Here $n_i$ is the *Boolean Term Occupancy Number* (BTON) as defined in Eq. (3.9c). A vector in Boolean Fock space represent a *concept*, therefore, it is also called ***Concept Fock Space*** (CFS) [4]. The $t$-dimensional basis for CFS-induced sample space $\Omega_F$ has following properties:

$$\langle \vec{n}_i \mid \vec{n}_j \rangle = \delta_{i,j} \qquad \sum_{i=1}^{t} |\vec{n}_i\rangle\langle \vec{n}_i| = \hat{I} \tag{4.3}$$

In CFS, the documents and the query in the GF-example (see App A) are represented as:

$$|d_1\rangle \to |\vec{d}_1\rangle = \sum_{i=1}^{t} n_{1,i} |\vec{n}_i\rangle = |1,0,1,0,1,1,1,1,1,0,0\rangle \tag{4.4a}$$

$$|d_2\rangle \to |\vec{d}_2\rangle = \sum_{i=1}^{t} n_{2,i} |\vec{n}_i\rangle = |1,1,0,1,0,0,1,1,0,1,1\rangle \tag{4.4b}$$

$$|d_3\rangle \to |\vec{d}_3\rangle = \sum_{i=1}^{t} n_{3,i} |\vec{n}_i\rangle = |1,1,0,0,0,1,1,1,1,0,1\rangle \tag{4.4c}$$

$$|q\rangle \to |\vec{q}\rangle = \sum_{i=1}^{t} n_{q,i} |\vec{n}_i\rangle = |0,0,0,0,0,1,0,0,0,1,1\rangle \tag{4.4d}$$

In some models (***Hiemstra*** [8]), it is assumed that the terms existing in documents but not in the query have no effect on RDQ; therefore, mapping Eq. (4.3) to single-term concepts, we have the following basis:

$$|\vec{n}_i\rangle \to |\vec{k}_i) = |0,..,k_i,...,0): \quad P(\vec{k}_i \,|\, \vec{k}_j) = \delta_{i,j}, \quad \sum\nolimits_i |\vec{k}_i)P(\vec{k}_i| = I \tag{4.5a}$$

Using map (4.5a), we have the following conditional probabilities:

$$P(\vec{k}_i \,|\, d_\mu) = P(k_i \,|\, d_\mu), \quad P(\vec{k}_i \,|\, q) = P(k_i \,|\, q) \tag{4.6a}$$

In other models [7, 9], the terms not in the query are also taken into account, and the map in (4.5a) is modified to:

$$|\vec{n}_i\rangle \to |\vec{k}_i) = |0,..,k_i,...,0)\prod\nolimits_{j\neq i}|\bar{k}_j): \quad P(\vec{k}_i \,|\, \vec{k}_j) = \delta_{i,j}, \quad \sum\nolimits_i |\vec{k}_i)P(\vec{k}_i| = I \tag{4.5b}$$

Using map (4.5b), we have the following conditional probabilities:

$$P(\vec{k}_i \,|\, d_\mu) = P(k_i \,|\, d_\mu)\prod\nolimits_{j\neq i}P(\bar{k}_j \,|\, d_\mu), \quad P(\vec{k}_i \,|\, q) = P(k_i \,|\, q)\prod\nolimits_{j\neq i}P(\bar{k}_j \,|\, q) \tag{4.6b}$$

Here we have used negation of a term:

$$\bar{k}_i \equiv \neg k_i \tag{4.7}$$

Now we derive formulas for some probabilistic models induced by CFS.

***Inference Network Model*** (INM, §2.8.2 of [5]): The ranking of a document $d_\mu$ with respect to a query $q$ is given by:

$$P(q \wedge d_\mu) \underset{(2.6)}{=} P(q \,|\, d_\mu)P(d_\mu) = P(q \,|\, I \,|\, d_\mu)P(d_\mu) \underset{(4.5)}{=} \sum\nolimits_{i=1}^{t} P(q \,|\, \vec{k}_i)P(\vec{k}_i \,|\, d_\mu)P(d_\mu) \tag{4.8}$$

Now let us apply Eq. (4.6b) into Eq. (4.8), we get:

$$P(q \wedge d_\mu) = \sum\nolimits_{i=1}^{t} P(q \,|\, \vec{k}_i)\left( P(k_i \,|\, d_\mu) \times \prod_{j\neq i} P(\bar{k}_j \,|\, d_\mu) \right)P(d_\mu) \tag{4.9}$$

This looks like the Eq. (2.7) of [5]. But, in our formula, $t$- dimensional basis from (4.2) is used, while in [5], the $2^t$-dimensional basis from Eq. (4.1) is used first, and then it is changed to basis similar to (4.5b) for ***tf-idf ranking strategies*** (pp. 54-55, [5]).

As in §3, we are interested in a symmetric formula. From *Bayes' formula* (2.7), we have:

$$P(q \,|\, \vec{k}_i) = \frac{P(\vec{k}_i \,|\, q)}{P(\vec{k}_i)}P(q) \tag{4.10}$$

Applying it to Eq. (4.8), we obtain:

$$P(q \wedge d_\mu) = P(q)P(d_\mu)\sum_{i=1}^{t} \frac{1}{P(\vec{k}_i)} P(\vec{k}_i \mid q)P(\vec{k}_i \mid q) \tag{4.11}$$

If we use expansion (4.6b) to Eq. (4.11), we have:

$$\begin{aligned} &P(q \wedge d_\mu) = P(q)P(d_\mu)\times \\ &\sum_{i=1}^{t} \frac{1}{P(k_i)} \left( P(k_i \mid q) \prod_{j\neq i} P(\bar{k}_j \mid q) \right) \left( \prod_{n_i=1} P(k_i \mid d_\mu) \prod_{j \neq i} P(\bar{k}_j \mid d_\mu) \right) \end{aligned} \tag{4.12}$$

Eq. (4.12) is an extension of Eq. (3.8). The extra factors can be calculated as follows:

$$\begin{aligned} P(\bar{k}_j \mid q) &= 1 - P(k_j \mid q) = 1 - w_{q,j}^2 \\ P(\bar{k}_j \mid d_\mu) &= 1 - P(k_j \mid d_\mu) = 1 - w_{\mu,j}^2 \end{aligned} \tag{4.13}$$

The *symmetric* ranking formula for CFS-INM now reads:

$$\text{INM: } P(q \wedge d_\mu) = C \cdot P(q)\, P(d_\mu) \sum_{i=1}^{t} \frac{w_{q,i}^2 \times w_{\mu,i}^2}{P(k_i)} \prod_{j \neq i} (1 - w_{q,j}^2) \times (1 - w_{\mu,j}^2) \tag{4.14}$$

If we use expansion (4.6a) (***Hiemstra*** [8]) to Eq. (4.8), then Eq. (4.12) is reduced to Eq. (3.8) and is evaluated just as in Eq. (3.11). This means that Eq. (4.11) and (4.14) of CFS-INM are natural extensions of Eq. (3.8) and (3.11) of TVS-INM.

For ***Belief Network Model*** (BNM, [5] and [8]), the RDQ formula becomes:

$$\text{BNM: } P(d_\mu \mid q) = \frac{P(d_\mu \wedge q)}{P(q)} \underset{(4.14)}{=} C \cdot P(d_\mu) \sum_{i=1}^{t} \frac{w_{q,i}^2 \times w_{\mu,i}^2}{P(k_i)} \prod_{j \neq i} (1 - w_{q,j}^2) \times (1 - w_{\mu,j}^2) \tag{4.15}$$

As mentioned in Eq. (3.12), Eq. (4.15) will produce the same result as Eq. (4.14).

Following ***P&C*** [9], the RDQ formula now reads

$$\text{P\&C: } P(q \mid d_\mu) = \frac{P(q \wedge d_\mu)}{P(d_\mu)} \underset{(4.14)}{=} C \cdot P(q) \sum_{i=1}^{t} \frac{w_{q,i}^2 \times w_{\mu,i}^2}{P(k_i)} \prod_{j \neq i} (1 - w_{q,j}^2) \times (1 - w_{\mu,j}^2) \tag{4.16}$$

Eq. (4.14) is symmetric and can be used to calculate RDD:

$$\text{INM: } P(d_\mu \wedge d_\nu) = C \cdot P(d_\mu)\, P(d_\nu) \sum_{i=1}^{t} \frac{w_{\mu,i}^2 \times w_{\nu,i}^2}{P(k_i)} \prod_{j \neq i} (1 - w_{\mu,j}^2) \times (1 - w_{\nu,j}^2) \tag{4.17}$$

According to Fuhr [10], in ***probabilistic concept space*** (***PCS***) model, a query reads:

$$\langle q|\mapsto\; P(q| = P(\bigcap_{i=1}^{t}\{[k_i]^{n_{q,i}} \wedge [\bar{k}_i]^{1-n_{q,i}}\}| \qquad (4.18)$$

For example, the query in our GF-example, Eq. (4.4d), now maps to:

$$P(q| = P(\bar{k}_1 \wedge \bar{k}_2 \wedge \bar{k}_3 \wedge \bar{k}_4 \wedge \bar{k}_5 \wedge k_6 \wedge \bar{k}_7 \wedge \bar{k}_8 \wedge \bar{k}_9 \wedge k_{10} \wedge k_{11}| \qquad (4.19)$$

Then the ranking formula for ***PCS-INM-Fuhr*** reads:

$$P(q \wedge d_\mu) = P(q|d_\mu)P(d_\mu) = P(d_\mu)\prod_{n_{q,i}=1} P(k_i|d_\mu)\prod_{n_{q,j}=0} P(\bar{k}_j|d_\mu) \qquad (4.20)$$

$$P(q \wedge d_\mu) = P(d_\mu)\prod_{i=1}^{t}(w_{\mu,i}^2)^{n_{q,i}}(1-w_{\mu,i}^2)^{1-n_{q,i}} \qquad (4.21)$$

Unfortunately, Eq. (4.20-21) may not be useful, because:

$$0 = P(q \wedge d_\mu),\ \text{if}\ \ \exists i: n_{q,i} \neq 0\ \text{but}\ w_{\mu,i} = 0 \qquad (4.22)$$

***TVS vs. CFS***: CFS is by nature a product of 2D-*factor spaces*. The occupancy of one term is independent of the occupancy of another term in a concept, so *different terms* are *independent events* in CFS, as in Eq. (2.21), although they are *disjoint events* in *Term Vector Space*, as in Eq. (3.4).

$$\text{In Term Vector Space:}\quad P(k_1|k_2) = 0 \quad \Rightarrow \quad P(k_1 \wedge k_2|\Omega_V) = 0 \qquad (4.23)$$

$$\text{In Concept Fock Space:}\quad P(\vec{k}_1 \wedge \vec{k}_2|\Omega_F) = P(\vec{k}_1|\Omega_F)P(\vec{k}_2|\Omega_F) \qquad (4.24)$$

Hence, we don't see the inconsistency raised by Fuhr in Ref [10].

Moreover, our CFS-induced models actually contain all the ingredients of the three classical IR models: the weights from vector space model (VSM), the conditional probabilities from Probabilistic models and CFS (or *Boolean Fock Space* [4]) from Boolean models. Therefore, armed with Dirac notation, PBN and CFS, we now have a platform to represent unified IR modeling.

## 5. Numerical Test Results Using Grossman-Frieder Example

The ***GF-Example*** ([6], or Table A.1 in Appendix A) has a query ($q$) and 3 documents ($\{d_1, d_1, d_1\}$). We now test our formulas using this example.

As already explained in the text, we will not use Eq. (3.12) and (4.15) for BIM models.

For comparison, results of Eq. (3.13) for VSM are listed. Because Eq. (3.13) does not have any unknown constant, we adjust all our models to have the same highest rank as given by VSM (the middle column in the following tables).

In addition, the results from *SVD metric model*, Eq. (3.3.4) of Ref. [4], are also listed. Using the metric tensor defined in Eq. (3.4.2) of [4], the relevance is given by:

$$\text{SVDM: } \cos(\theta_{\alpha,\beta}) = \langle \upsilon_\alpha | \hat{g} | \upsilon_\beta \rangle = \frac{\langle d_\alpha | \hat{g} | d_\beta \rangle}{\sqrt{\langle d_\alpha | \hat{g} | d_\alpha \rangle \langle d_\beta | \hat{g} | d_\beta \rangle}} \quad (5.1)$$

Note there is no adjustable constant in Eq. (5.1).

There are two test cases, the first uses Eq. (3.9) and the second uses (3.10). Each case will use four different weight formulas, given by Eq. (3.6), (3.19), (3.20) and (3.21), listed in Table B.1-4 in Appendix B, respectively.

## 5.1-Test case I: Using APDQK-I or Eq. (3.9)

APDQK-I, or Eq. (3.9), evaluates absolute probabilities as follows:

$$P(q) \propto \frac{1}{\sum_{i=1}^{t} tf_{q,i}}, \quad P(q) \propto \frac{1}{\sum_{i=1}^{t} tf_{q,i}}, \quad P(k_i) \propto \frac{N_i + 1}{N + 1} \quad (5.2)$$

**Case I-1:** ***Test results with WF-1 defined in Eq. (3.6)***

**Table 5.1A:** The Relevance of Documents Related to the Query (weights in Table B.1)

| Formula used | $d_1$ | $d_2$ | $d_3$ |
|---|---|---|---|
| Eq. (3.11): TVS-INM, *C*=169 | 0.1277 | 0.5477 | 0.2555 |
| Eq. (3.13): TVS-P&C, *C*=21.1 | 0.1118 | 0.5477 | 0.2235 |
| Eq. (4.14): CFS-INM, C=751 | 0.1001 | 0.5477 | 0.2002 |
| **Eq. (4.16): CFS-P&C, C=10.6** | **0.0876** | **0.5477** | **0.1751** |
| Eq. (3.15): VSM | 0.2182 | 0.5477 | 0.4364 |
| Eq. (3.3.4) of [4]: SVDM | -0.0552 | 0.9912 | 0.4480 |

**Table 5.1B:** The Closeness (Relevance Ranking) of Documents (weights in Table B.1)

| Formula used | $d_1 \leftrightarrow d_2$ | $d_1 \leftrightarrow d_3$ | $d_2 \leftrightarrow d_3$ |
|---|---|---|---|
| Eq. (3.14): TVS-INM, *C*=519 | 0.2647 | 0.7143 | 0.4375 |
| Eq. (4.17): CFS-INM, C=7698 | 0.2072 | 0.7143 | 0.3901 |
| Eq. (3.16): VSM | 0.3585 | 0.7143 | 0.5976 |
| **Eq. (5.1): SVDM** | **-0.1892** | **0.8678** | **0.3228** |

We see that all formulas produce consistent and comparable results; Except Eq. (3.22), they all give the same order of relevancies, although the relative ratio changes:

RDQ: $R(d_2,q) > R(d_3,q) > R(d_1,q)$ (5.3)

RDD: $R(d_1,d_3) > R(d_2,d_3) > R(d_1,d_2)$ (5.4)

Note that the ranking order in Eq. (5.3) is consistent with most results obtained for the same example using various IR models in Ref [6] (see Table C.1 & C.2).

If we believe that the better the model if the bigger the difference in relevancies, then we see that the efficiency of models is in following order:

RDQ: $CFS > SVDM > TVS > VSM$ (5.5a)

RDD: $SVDM > CFS > TVS > VSM$ (5.5b)

**Case I-2:** ***Test results with WF-2 defined in Eq. (3.19)***.

**Table 5.2A:** The Relevance of Documents Related to the Query (weights in Table B.2)

| Formula used | $d_1$ | $d_2$ | $d_3$ |
|---|---|---|---|
| Eq. (3.11): TVS-INM, *C*=66.19 | 0.0067 | 0.8249 | 0.0562 |
| Eq. (3.13): TVS-P&C, *C*=12.63 | 0.0059 | 0.8249 | 0.0492 |
| Eq. (4.14): CFS-INM, C=234.0 | 0.0006213 | 0.8249 | 0.0209 |
| **Eq. (4.16): CFS-P&C, C=30.00** | **0.0005436** | **0.8249** | **0.0074** |
| Eq. (3.15): VSM | 0.0801 | 0.8249 | 0.3272 |
| Eq. (3.3.4) of [4]: SVDM | -0.0552 | 0.9912 | 0.4480 |

**Table 5.2B:** The Closeness (Relevance Ranking) of Documents (weights in Table B.2)

| Formula used | $d_1 \leftrightarrow d_2$ | $d_1 \leftrightarrow d_3$ | $d_2 \leftrightarrow d_3$ |
|---|---|---|---|
| Eq. (3.14): TVS-INM, *C*=534 | 0.0 | 0.2448 | 0.0923 |
| **Eq. (4.17): CFS-INM, C=9986** | **0.0** | **0.2448** | **0.0596** |
| Eq. (3.16): VSM | 0.0 | 0.2448 | 0.1607 |
| Eq. (5.1): SVDM | -0.1892 | 0.8678 | 0.3228 |

We see all models give the same ranking order as in Eq. (5.3-4) and their efficiency order for both RDQ and RDD are:

$CFS > TVS > SVDM > VSM$ (5.6)

**Case I-3:** ***Test results with WF-3 defined in Eq.*** **(*3.20*)**. We see all models, except Eq. (3.22), give the same ranking order as in Eq. (5.3-4) and they have the same efficiency order as Eq. (5.6).

**Table 5.3A:** The Relevance of Documents Related to the Query (weights in Table B.3)

| Formula used | $d_1$ | $d_2$ | $d_3$ |
|---|---|---|---|
| Eq. (3.11): TVS-INM, *C*=121.4 | 0.0385 | 0.5799 | 0.3212 |
| Eq. (3.13): TVS-P&C, *C*=15.17 | 0.0337 | 0.5799 | 0.2810 |
| Eq. (4.14): CFS-INM, C=408.8 | 0.0170 | 0.5799 | 0.2028 |
| **Eq. (4.16): CFS-P&C, C=30.00** | **0.0149** | **0.5799** | **0.1774** |
| Eq. (3.15): VSM | 0.1413 | 0.5799 | 0.5773 |
| Eq. (3.3.4) of [4]: SVDM | -0.0552 | 0.9912 | 0.4480 |

**Table 5.3B:** The Closeness (Relevance Ranking) of Documents (weights in Table B.3)

| Formula used | $d_1 \leftrightarrow d_2$ | $d_1 \leftrightarrow d_3$ | $d_2 \leftrightarrow d_3$ |
|---|---|---|---|
| Eq. (3.14): TVS-INM, *C*=534 | 0.0 | 0.2448 | 0.1286 |
| **Eq. (4.17): CFS-INM, C=9986** | **0.0** | **0.2448** | **0.1040** |
| Eq. (3.16): VSM | 0.0 | 0.2448 | 0.1897 |
| Eq. (5.1): SVDM | -0.1892 | 0.8678 | 0.3228 |

**Case I-4:** ***Test results with WF-4 defined in Eq. (3.21)***. This time, for VMS and all our models, the order of relevance of document to query is changed to:

RDQ: $R(d_2,q) > R(d_1,q) > R(d_2,q)$ (5.7)

It shows that *weight formula* (3.21) *may not be a good choice for RDQ*.

**Table 5.4A:** The Relevance of Documents Related to the Query (weights in Table B.4)

| Formula used | $d_1$ | $d_2$ | $d_3$ |
|---|---|---|---|
| Eq. (3.11): TVS-INM, *C*=101.1 | 0.2610 | 0.8146 | 0.0653 |
| Eq. (3.13): TVS-P&C, *C*=12.63 | 0.2284 | 0.8146 | 0.0571 |
| Eq. (4.14): CFS-INM, C=234.0 | 0.1035 | 0.8146 | 0.0209 |
| Eq. (4.16): CFS-P&C, C=30.00 | 0.0905 | 0.8146 | 0.0183 |
| Eq. (3.15): VSM | 0.5525 | 0.8146 | 0.3829 |

But *weight formula* (3.21) gives the same RDD order as in Eq. (5.4).

**Table 5.4B:** The Closeness (Relevance Ranking) of Documents (weights in Table B.4)

| Formula used | $d_1 \leftrightarrow d_2$ | $d_1 \leftrightarrow d_3$ | $d_2 \leftrightarrow d_3$ |
|---|---|---|---|
| Eq. (3.14): TVS-INM, *C*=534 | 0.0 | 0.2448 | 0.0923 |
| **Eq. (4.17): CFS-INM, C=9986** | **0.0** | **0.2448** | **0.0583** |
| Eq. (3.16): VSM | 0.0 | 0.2448 | 0.1607 |

From the above test results, we can conclude that:

1. WF-2, Eq. (3.19) is our best weight formula for VMS and our models.
2. CFS- P&C model, Eq. (4.16), is the best for RDQ.
3. CFS- INM model, Eq. (4.15), is the best for RDD.
4. WF-4, Eq. (3.21) produces inconsistent results of RDQ for VMS and our models.

## 5.2-Test case II: Using APDQK-II or Eq. (3.10)

APDQK-II, or Eq. (3.10), evaluates absolute probabilities as follows:

$$P(q)=\frac{|q|}{|\Omega_V|}=\frac{\sum_{i=1}^{t} n_{q,i}}{t},\quad P(d_\mu)=\frac{|d_\mu|}{|\Omega_V|}=\frac{\sum_{i=1}^{t} n_{\mu,i}}{t},\quad P(k_i)=\frac{|k_i|}{|\Omega_V|}=\frac{1}{t} \tag{5.8}$$

We will not use Eq. (3.13) and (4.16) for P&C models, since they will give relative ranking identical to INM models, due to fact that all three documents happen to have the same absolute probability based on Eq. (5.8):

$$P(d_1)=P(d_2)=P(d_3)=7/11 \tag{5.9}$$

Comparing all test results in Test Case II, we can make the same conclusion as for Test Case I, except there is no difference between P&C and INM when ranking RDQ. Hence we need only show the results for its best scenario (Case II-2):

**Case II-2:** ***Test results with weights defined in Eq. (3.19)***.

**Table 5.5A:** The Relevance of Documents Related to the Query (weights in Table B.2)

| Formula used | $d_1$ | $d_2$ | $d_3$ |
|---|---|---|---|
| Eq. (3.11): TVS-INM, *C*=0.005001 | 0.008835 | 0.8248 | 0.07368 |
| **Eq. (4.14): CFS-INM, C=1.1813** | **0.0008155** | **0.8248** | **0.009736** |
| Eq. (3.15): VSM | 0.0801 | 0.8248 | 0.3272 |

**Table 5.5B:** The Closeness (Relevance Ranking) of Documents (weights in Table B.2)

| Formula used | $d_1 \leftrightarrow d_2$ | $d_1 \leftrightarrow d_3$ | $d_2 \leftrightarrow d_3$ |
|---|---|---|---|
| Eq. (3.14): TVS-INM, *C*=1.8338 | 0.0 | 0.2448 | 0.1055 |
| **Eq. (4.17): CFS-INM, C=34.285** | **0.0** | **0.2448** | **0.1055** |
| Eq. (3.16): VSM | 0.0 | 0.2448 | 0.1607 |

Compare Case II-2 with Case I-2, we see that their results do not have any significant difference. Therefore, $P(d_\mu)$, $P(q)$ and $P(k_i)$ can be estimated either from APDQK-1, Eq. (3.9), or from APDQK-2, Eq. (3.10).

Our RDD ranking results are consistent with Diffusion map [12-13]. For t > 0, the distance between the three documents are in the same order (see Eq. (3.4.12) of [14]):

$$D_t^2(1,2) > D_t^2(2,3) > D_t^2(1,3) \tag{5.10}$$

Our RDQ raking results are also consistent with almost all the results given by Ref. [6]. The best outcomes in Ref [6] are from VMS model (bold row in Table C.2), but the results from our CFS-induced models using weight formula Eq. (3.19) seem to be better (see Table 5.2A or 5.4A). The worst outcomes in Ref [6] are from INM model, which are inconsistent with other results, as shown in Eq. (C.2) or the red row in Table C.2.

## 6. Summary

In this paper, we exposed the close relations between Dirac notation and our *PBN* in time-independent systems with discrete observable:

$$|\,\Omega) \leftrightarrow |\,\psi\rangle, \quad P(x_i) = P(x_i\,|\,\Omega) = |\langle x_i\,|\,\psi\rangle\,|^2 \tag{6.1}$$

$$1 = P(\Omega\,|\,\Omega) \leftrightarrow \langle\psi\,|\,\psi\rangle = \sum_i |\,\langle x_i\,|\,\psi\rangle\,|^2 = 1 \tag{6.2}$$

Applying to *Term Vector Space* (TVS), we obtained probability distribution functions for documents and query, based on their term weights as used in classical VSM.

Next, we discussed various probabilistic models induced by TVS and by *Concept Fock Space* (CFS) and derived their expressions of RDQ and/or RDD. Then we tested our expressions by applying them to the famous *GF*-example with various scenarios based on two APDQ and four weight formulas (WF). Our test results are consistent with each other and with other models.

The CFS-induced models contain the ingredients of *all three classic IR models*. Hence, by combining Dirac notation, PBN and CFS, we now may have a platform to develop unified IR models.

Of cause, our ranking formulas derived for induced probabilistic models need to be tested against bigger or real data sets. We also need more examination on our proposals about *PBN* [3], which can be extended to continuous variables (like positions) and to time dependent variables (like Markov chains).

## Appendix A: The Grossman-Frieder example

In reference [6], a simple example is used throughout that book. The example, referred to *GF-Example* in this article, has a collection of three Documents and one Query as follows:

$q$ : "*gold silver truck*"
$d_1$: "Shipment of gold damaged in a fire"
$d_2$: "Delivery of silver arrived in a silver truck"
$d_3$: "Shipment of gold arrived in a truck"

The basic data are presented in table A.1.

**Table A.1**: Term Frequencies ($tf_{\mu,i}, tf_{q,i}$) and ($N_i, idf_i$) of GF-Example

| Term $i$ | 1 | 2 | 3 | 4 | 5 | 6 | 7 | 8 | 9 | 10 | 11 |
|---|---|---|---|---|---|---|---|---|---|---|---|
| Word | a | arrived | damaged | delivery | fire | ***gold*** | in | of | shipment | ***silver*** | ***truck*** |
| $tf_{1,i}$ | 1 | 0 | 1 | 0 | 1 | 1 | 1 | 1 | 1 | 0 | 0 |
| $tf_{2,i}$ | 1 | 1 | 0 | 1 | 0 | 0 | 1 | 1 | 0 | 2 | 1 |
| $tf_{3,i}$ | 1 | 1 | 0 | 0 | 0 | 1 | 1 | 1 | 1 | 0 | 1 |
| $tf_{q,i}$ | 0 | 0 | 0 | 0 | 0 | 1 | 0 | 0 | 0 | 1 | 1 |
| $N_i$ | 3 | 2 | 1 | 1 | 1 | 2 | 3 | 3 | 2 | 1 | 2 |
| $idf_i$ | 0 | 0.176 | 0.477 | 0.477 | 0.477 | 0.176 | 0 | 0 | 0.176 | 0.477 | 0.176 |

## Appendix B: Weight Formulas (WF) and Weights of the GF- Example

**Table B.1**: Term Weights ($w_{\mu,i}, w_{q,i}$), based on WF-1, Eq. (3.6):

$$\text{WF-1:}\quad w_{\mu,i} = \frac{tf_{\mu,i}}{\sqrt{\sum_{j=1}^{t} tf_{\mu,j}^2}}, \qquad w_{q,i} = \frac{tf_{q,i}}{\sqrt{\sum_{j=1}^{t} tf_{\mu,j}^2}} \tag{B.1}$$

| Term $i$ | 1 | 2 | 3 | 4 | 5 | 6 | 7 | 8 | 9 | 10 | 11 |
|---|---|---|---|---|---|---|---|---|---|---|---|
| Word | a | arrived | damaged | delivery | fire | ***gold*** | in | of | shipment | ***silver*** | ***truck*** |
| $w_{1,i}$ | 0.378 | 0 | 0.378 | 0 | 0.378 | 0.378 | 0.378 | 0.378 | 0.378 | 0 | 0 |
| $w_{2,i}$ | 0.316 | 0.316 | 0 | 0.316 | 0 | 0 | 0.316 | 0.316 | 0 | 0.632 | 0.316 |
| $w_{3,i}$ | 0.378 | 0.378 | 0 | 0 | 0 | 0.378 | 0.378 | 0.378 | 0.378 | 0 | 0.378 |
| $w_{q,i}$ | 0 | 0 | 0 | 0 | 0 | 0.577 | 0 | 0 | 0 | 0.577 | 0.577 |

**Table B.2**: Term Weights ($w_{\mu,i}, w_{q,i}$), based on WF-2, Eq. (3.19):

WF-2: $w_{\mu,i} = \dfrac{tf_{\mu,i} \cdot idf_i}{\sqrt{\sum_{i=1}^{t}\left(tf_{\mu,i} \cdot idf_i\right)^2}}, \quad w_{q,i} = \dfrac{tf_{q,i} \cdot idf_i}{\sqrt{\sum_{i=1}^{t}\left(tf_{\mu,i} \cdot idf_i\right)^2}}$ (B.2)

| Term $i$ | 1 | 2 | 3 | 4 | 5 | 6 | 7 | 8 | 9 | 10 | 11 |
|---|---|---|---|---|---|---|---|---|---|---|---|
| $w_{1,i}$ | 0 | 0 | 0.663 | 0 | 0.663 | 0.245 | 0 | 0 | 0.245 | 0 | 0 |
| $w_{2,i}$ | 0 | 0.161 | 0 | 0.435 | 0 | 0 | 0 | 0 | 0 | 0.871 | 0.161 |
| $w_{3,i}$ | 0 | 0.500 | 0 | 0 | 0 | 0.500 | 0 | 0 | 0.500 | 0 | 0.500 |
| $w_{q,i}$ | 0 | 0 | 0 | 0 | 0 | 0.327 | 0 | 0 | 0 | 0.823 | 0.327 |

**Table B.3**: Term Weights ( $w_{\mu,i}$ , $w_{q,i}$ ), based on WF-3, Eq. (3.20):

WF-3: $w_{\mu,i} = \dfrac{\log(1+tf_{\mu,i}) \cdot idf_i}{\sqrt{\sum_{i=1}^{t}\left[\log(1+tf_{\mu,i}) \cdot idf_i\right]^2}}, \quad w_{q,i} = \dfrac{\log(1+tf_{q,i})}{\sqrt{\sum_{i=1}^{t}\left[\log(1+tf_{\mu,i}) \cdot idf_i\right]^2}}$ (B.3)

| Term $i$ | 1 | 2 | 3 | 4 | 5 | 6 | 7 | 8 | 9 | 10 | 11 |
|---|---|---|---|---|---|---|---|---|---|---|---|
| $w_{1,i}$ | 0 | 0 | 0.663 | 0 | 0.663 | 0.245 | 0 | 0 | 0.245 | 0 | 0 |
| $w_{2,i}$ | 0 | 0.190 | 0 | 0.514 | 0 | 0 | 0 | 0 | 0 | 0.815 | 0.190 |
| $w_{3,i}$ | 0 | 0.500 | 0 | 0 | 0 | 0.500 | 0 | 0 | 0.500 | 0 | 0.500 |
| $w_{q,i}$ | 0 | 0 | 0 | 0 | 0 | 0.577 | 0 | 0 | 0 | 0.577 | 0.577 |

**Table B.4**: Term Weights ( $w_{\mu,i}$ , $w_{q,i}$ ), based on WF-4, Eq. (3.21):

WF-4: $w_{\mu,i} = \dfrac{f_{\mu,i} \cdot idf}{\sqrt{\sum_{i=1}^{t}(f_{\mu,i} \cdot idf_i)^2}}, \quad w_{q,i} = \dfrac{\left(1+f_{q,i}\right) \cdot idf}{\sqrt{\sum_{i=1}^{t}\left((1+f_{q,i}) \cdot idf_i\right)^2}}$ (B.4)

| Term $i$ | 1 | 2 | 3 | 4 | 5 | 6 | 7 | 8 | 9 | 10 | 11 |
|---|---|---|---|---|---|---|---|---|---|---|---|
| $idf_i$ | 0 | 0.176 | 0.477 | 0.477 | 0.477 | 0.176 | 0 | 0 | 0.176 | 0.477 | 0.176 |
| $w_{1,i}$ | 0 | 0 | 0.663 | 0 | 0.663 | 0.245 | 0 | 0 | 0.245 | 0 | 0 |
| $w_{2,i}$ | 0 | 0.161 | 0 | 0.435 | 0 | 0 | 0 | 0 | 0 | 0.871 | 0.161 |
| $w_{3,i}$ | 0 | 0.500 | 0 | 0 | 0 | 0.500 | 0 | 0 | 0.500 | 0 | 0.500 |
| $w_{q,i}$ | 0 | 0.128 | 0.346 | 0.346 | 0.346 | 0.255 | 0 | 0 | 0.128 | 0.692 | 0.255 |

## Appendix C: Test Results of the Grossman-Frieder Example in Ref. [6]

In Table C.1 we have listed the test results of some models in Ref [6]. The weight formula for VMS on page 16 of [6] is similar to WF-2, Eq. (3.19), but not normalized:

$$w_{\mu,i} = tf_{\mu,i} \cdot idf, \quad w_{q,i} = tf_{q,i} \cdot idf_i \tag{C.1}$$

**Table C.1:** The Relevance of Documents Related to the Query

| **Result quoted from Ref.** [6] | $d_1$ | $d_2$ | $d_3$ |
|---|---|---|---|
| VMS, page 16 | 0.031 | 0.486 | 0.062 |
| Poisson Model, page 40, no *tf* | -0.477 | 1.653 | 0.699 |
| Poisson Model, page 40, with *tf* | -0.484 | 2.269 | 0.708 |
| Language Model, page 51 | 0.000409 | 0.00121 | 0.000743 |
| Dirichlet Priors, page 55 | 0.0000114 | 0.0002590 | 0.0001728 |
| Jelinek-Mercer, page 56 | 0.000314 | 0.000443 | 0.000381 |
| Absolute Discount, page 56 | 0.001215 | 0.021716 | 0.005727 |
| INM, page 66-67 | 0.237 | 0.473 | 0.511 |
| LSI, page 73 | -0.0541 | 0.9910 | 0.9543 |

To better compare the results, in Table C.2, the highest relevancies (in middle column) are adjusted to the VMS value (0.8249) from Table 5.2A. One can see that the RDQ *result of INM from* [6] (in red) *is inconsistent with other models*:

$$\text{RDQ: } R(d_3,q) > R(d_2,q) > R(d_1,q) \tag{C.2}$$

**Table C.2:** The Adjusted Relevance of Documents Related to the Query

| Result quoted from Ref. [6] | $d_1$ | $d_2$ | $d_3$ |
|---|---|---|---|
| **VMS, page 16** | **0.0526** | **0.8249** | **0.1052** |
| Poisson Model, page 40, no *tf* | -0.2380 | 0.8249 | 0.3488 |
| Poisson Model, page 40, with *tf* | -0.1760 | 0.8249 | 0.2574 |
| Language Model, page 51 | 0.2788 | 0.8249 | 0.5065 |
| Dirichlet Priors, page 55 | 0.0363 | 0.8249 | 0.5504 |
| Jelinek-Mercer, page 56 | 0.5847 | 0.8249 | 0.7095 |
| Absolute Discount, page 56 | 0.0461 | 0.8249 | 0.2175 |
| INM, page 66-67 | 0.4133 | 0.8249 | 0.8912 |
| LSI, page 73 | -0.0450 | 0.8249 | 0.7944 |